\documentclass[twocolumn,showpacs,fleqn,nobibnotes]{revtex4}

\usepackage{amsmath}
\usepackage{graphicx}
\usepackage{float}
\usepackage{subfigure}

\def\lsim{\raise0.3ex\hbox{$<$\kern-0.75em\raise-1.1ex\hbox{$\sim$}}}
\def\gsim{\raise0.3ex\hbox{$>$\kern-0.75em\raise-1.1ex\hbox{$\sim$}}}

\begin{document}

\title{Exclusive photoproduction of quarkonium in proton-nucleus collisions at the energies available at the CERN Large Hadron Collider}
\pacs{12.38.Bx; 13.60.Hb}
\author{G. Sampaio dos Santos and M.V.T. Machado}

\affiliation{High Energy Physics Phenomenology Group, GFPAE  IF-UFRGS \\
Caixa Postal 15051, CEP 91501-970, Porto Alegre, RS, Brazil}

\begin{abstract}
In this work we investigate the coherent photoproduction of $\psi(1S)$, $\psi(2S)$ and $\Upsilon$ states in the proton-nucleus collisions in the  LHC energies. Predictions for the rapidity distributions are presented  using the color dipole formalism and including saturation effects that are expected to be relevant at high energies. Calculations are done at the energy  5.02 TeV and also for the next LHC run at 8.8 TeV in proton-lead mode. Discussion is performed on the main theoretical uncertainties associated to the calculations.

\end{abstract}

\maketitle

\section{Introduction} 

The exclusive quarkonium photoproduction has being investigated both
experimentally and theoretically in recent years as it allows to test
perturbative Quantum Chromodynamics. The  masses of these heavy
mesons, $m_V$, give a perturbative scale for the problem even in the
photoproduction limit, $Q^2\rightarrow 0$.  An important feature of these exclusive
processes at the high energy regime is the possibility to investigate
the hard pQCD Pomeron exchange. For this energy domain hadrons and photons can
be considered as color dipoles in the mixed light cone representation,
where their transverse size can be considered frozen during the
interaction \cite{nik}. Therefore, the scattering process is characterized by the
color dipole cross section describing the interaction of those color
dipoles with the nucleon or nucleus target. Such  an approach is intuitive
and allows to introduce information on dynamics beyond the leading
logarithmic QCD approach. The information of the meson formation is given by their wavefunctions and to compute predictions for their excited states  is a reasonably easy task\cite{Nemchik:1996pp}.

In the present work, we investigate the exclusive production of
$J/\psi$, its radially excited $\psi (2S)$ state and $\Upsilon (1S)$
in proton-nucleus collisions in the LHC energy range.  The
theoretical framework considered is the light-cone dipole formalism \cite{nik},
where the $Q\bar{Q}$ fluctuation (color dipole) of the incoming
quasi-real photon (from the protons or nuclei) interacts with the target via the
dipole cross section and the result is projected in the wavefunction
of the observed hadron. At high energies, the transition of the regime
described by the linear dynamics of emissions chain to a new regime
where the physical process of recombination of partons becomes
important is expected. It is characterized by the limitation on the
maximum phase-space parton density that can be reached in the hadron
wavefunction, the  so-called parton saturation phenomenon (see reviews in Ref. \cite{hdqcd}). The
transition is set by saturation scale $Q_{\mathrm{sat}}\propto
x^{\lambda}$, which is enhanced in the nuclear case. We will make use of this formalism to evaluate the corresponding cross sections. In the $pA$
collisions considered here, it is possible to investigate  at the same
time the dynamics on the photon-proton cross section and on the
photon-nucleus cross section. The dominant contribution comes from the photon-proton interaction as the photon flux due to the nucleus is higher compared to that due to the proton. It will be shown that the photon-nucleus contribution is relevant at large rapidities and increasingly important for heavy quarkonia as the $\Upsilon$ states. The paper is organized as follows. In next section we summarize the main theoretical information to compute the rapidity distribution of quarkonia in $pA$ collisions. In section \ref{discuss} we present the numerical calculations and discuss the main theoretical uncertainties and a comparison with another approaches is done. In last section we show the main conclusions.

\section{Theoretical framework and main expressions} 

Lets consider the proton-nucleus interaction at large impact parameter ($b > R_p + R_A$) and at ultrarelativistic energies. In this regime we expect the electromagnetic interaction to be dominant. In this case, the cross section of quarkonium $V$ production can be evaluated within the Weizs\"{a}cker-Williams approximation as a product of the photon flux emitted by one of the colliding participants ad the cross section of quarkonium photoproduction on the remaining hadron or nucleus. In particular, in proton-lead collisions if the quarkonium rapidity, $y$, is positive in the nucleus beam direction its rapidity distribution reads as \cite{upcs}:
\begin{eqnarray}
\frac{d\sigma }{dy} (Pb+p\rightarrow Pb+p+V) & = & \frac{dN_{\gamma}^{Pb}(y)}{d\omega }\sigma_{\gamma p \rightarrow V +p}(y) \nonumber \\
& + &\frac{dN_{\gamma}^p(-y)}{d\omega }\sigma_{\gamma Pb \rightarrow V +Pb}(-y),\nonumber
\end{eqnarray}
where $\frac{dN_{\gamma}(y)}{d\omega }$ is the corresponding photon flux and $y = \ln (2\omega/m_V)$, with $\omega$ being the photon energy. The case for the inverse beam direction is straightforward. We use the Weisz\"acker-Williams method to calculate the flux of photons  from a charge $Z$ nucleus \cite{upcs}:
\begin{eqnarray}
\frac{dN_{\gamma}^{Pb}}{d\omega}  = \frac{2\,Z^2\alpha_{em}}{\pi}\left[\xi K_0\,(\xi )K_1\,(\xi)+ \frac{\xi^2}{2}\left(K_1^2\,(\xi )-  K_0^2\,(\xi) \right)\right],\nonumber
\label{fluxint1}
\end{eqnarray}
where $\xi=\omega\,(R_p + R_A)/\gamma_L$ and  $\gamma_L$ is the Lorentz boost  of a single beam. For the proton case, we use the following expression for the photon energy spectrum \cite{KN}
\begin{eqnarray}
\frac{dN_{\gamma}^p}{d\omega}  & =  & \frac{\alpha_{em}}{2\pi}\, \left[1+\left(1-\frac{2\omega}{\sqrt{s}} \right)^2\right] \nonumber \\
& \times & \left(\ln \chi - \frac{11}{6} + \frac{3}{\chi}-\frac{3}{2\chi^2}+\frac{1}{3\chi^3}\right), \nonumber
\label{fluxint2}
\end{eqnarray}
where $\chi = 1+ (Q_0^2/Q_{\mathrm{min}}^2)$ with $Q_0^2=0.71$ GeV$^2$ and $Q_{\mathrm{min}}^2=\omega^2/\gamma_L^2$.

In the present analysis, we consider the photon-hadron/nucleus scattering in the color dipole formalism, in which most of the energy is
carried by the hadron and the  photon  dissociates into a quark-antiquark pair long before the scattering. Such an approach is turns out the analysis of the small-$x$ dynamics of the hadron wavefunction more clear and intuitive.  The probing
projectile fluctuates into a
quark-antiquark pair with transverse separation $r$ long after the interaction, which then
scatters off the target (proton or nucleus) \cite{nik}. Accordingly, the cross section for exclusive photoproduction of quarkonium off a nucleon target is given by \cite{Nemchik:1996pp},
\begin{eqnarray}
\sigma_{\gamma p\rightarrow Vp}  =  \frac{1}{16\pi B_V} \left|\sum_{h, \bar{h}} \int dz\, d^2r \,\Psi^\gamma_{h, \bar{h}}\sigma_{dip}(x,r)\, \Psi^{V*}_{h, \bar{h}}  \right|^2, \nonumber
\label{sigmatot}
\end{eqnarray}
where $\Psi^{\gamma}$ and $\Psi^{V}$ are the light-cone wavefunction  of the photon  and of the  vector meson, respectively.  The Bjorken variable is denoted by $x$, the dipole cross section by  $\sigma_{dip}(x,r)$ and the  diffractive slope parameter by $B_V$.  Here, we consider the energy dependence of the slope using a Regge motivated expression \cite{H1psi2}. In our numerical calculation, corrections for skewedness and real part of amplitude have been also included \cite{GGM}. On the other hand, the exclusive photoproduction off nuclei for coherent processes can be computed in a simple way in the large coherence length $\ell_c\gg R_A$ limit \cite{Boris,Borispsi}:
\begin{eqnarray}
\sigma_{\gamma A \rightarrow VA} & = & \int d^2b\,|S(x,b;\,A)|^2 ,\nonumber \\
S(x,b;\,A) & = & \sum_{h, \bar{h}} \int dz\, d^2r \,\Psi^{\gamma}_{h, \bar{h}} (z,r)\Psi^{V*}_{h, \bar{h}}(z,r,m_V) \nonumber \\
&\times & \left[1-\exp\left(-\frac{1}{2}R_G\sigma_{dip}(x,r) T_A(b)\right) \right],\nonumber
 \label{eq:coher} 
\end{eqnarray} 
where $T_A(b)= \int dz\rho_A(b,z)$  is the nuclear thickness function. In the numerical evaluations, we have considered the boosted Gaussian wavefunction \cite{Sandapen} and the  phenomenological saturation model proposed in Ref. \cite{IIM} which encodes the main properties of the saturation approaches. We call attention that the parameters of model \cite{IIM} have been updated with the recent high precision combined data from HERA in Ref. \cite{AmirSchmidt}. Large $x$ effects have been introduced by multiplying the dipole cross section by  a factor $(1-x)^\alpha$. The nuclear ratio for the gluon density is denoted by $R_G(x,Q^2,b)$, which will be discussed in the following section.

\section{Results and discussions}
\label{discuss}
\begin{figure}[t]
\includegraphics[scale=0.47]{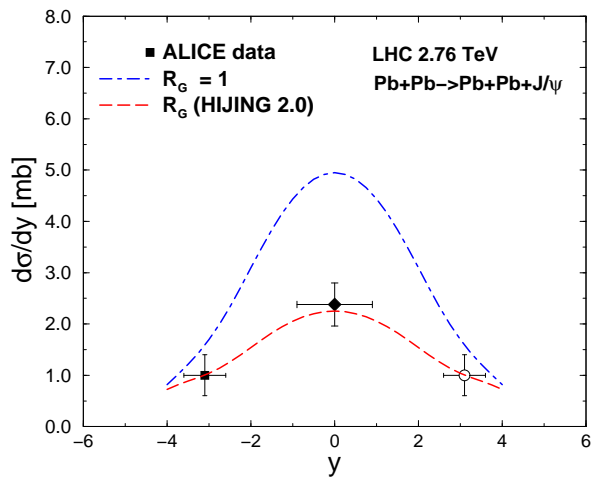}
\caption{(Color online) Rapidity distribution of $J/\psi$ photoproduction in PbPb collisions at 2.76 TeV compared to ALICE data\cite{ALICE1,ALICE2}. The dot-dashed curve represents the calculation using $R_G=1$ and the long-dashed one is for the $\mathrm{HIJING}$ 2.0 nuclear ratio.}
\label{fig:1}
\end{figure}
 
Before presenting the predictions for the quarkonia production let us discuss the role played by the gluon shadowing in the input cross section $\sigma_{\gamma A\rightarrow VA}$. It was recently shown that the color dipole approach with $R_G=1$ is not able to describe the ALICE data for coherent production of $J/\psi$ at 2.76 GeV \cite{ALICE1,ALICE2}. The mid-rapidity  cross section is overestimated by a factor two. The reason is that for $R_G=1$ in the photon-nucleus cross section the nuclear effect included via  eikonalization corresponds to lowest $Q\bar{Q}$ Fock component, $|Q\bar{Q}\rangle$. It does not include any correction for gluon shadowing, but rather correspond to shadowing of sea quarks in nuclei. Although $\sigma_{dip}$ includes all possibles effects of gluon radiation, the eikonal assumes that none of radiated gluons take part in multiple interactions in the nucleus. The leading order correction corresponding to gluon shadowing comes from the eikonalization of the next Fock component $|Q\bar{Q}G \rangle$. Explicitly, in the large coherence length limit $\ell_c \rightarrow \infty$ the general formula for the $\gamma A$ cross section is given by,
\begin{eqnarray}
\frac{d^2\sigma_{\gamma A}}{d^2b}& = & 2\int dz \int d^2r \left|\psi_{q\bar{q}}\right|^2\left[1-\exp \left(-\frac{1}{2} \sigma_{q\bar{q}}  T_A(b)\right) \right] \nonumber \\
 & + & 2 \int dz \int \frac{dz_G}{z_G} \int d^2r_1 \int d^2r_2  \left|\psi_{q\bar{q}G}\right|^2 \nonumber \\
& \times & \left[1-\exp \left(-\frac{1}{2} \sigma_{q\bar{q}G}  T_A(b)\right) \right], 
\end{eqnarray}
where $\psi_{q\bar{q}G}$ is the wavefunction for the gluonic component and the cross section for the three body system $\sigma_{q\bar{q}G}$  can be expressed in terms of the dipole cross section, $\sigma_{dip}$. The complete calculation has been done in Ref. \cite{KST}, and it was shown that the $|Q\bar{Q}G \rangle$ contribution can be absorbed in a factor $R_G(x,Q^2,b)$ multiplying the original dipole cross section. The evaluation of gluon shadowing contribution in \cite{KST} is not trivial and somewhat complex for practical use.

\begin{figure}[t]
\includegraphics[scale=0.47]{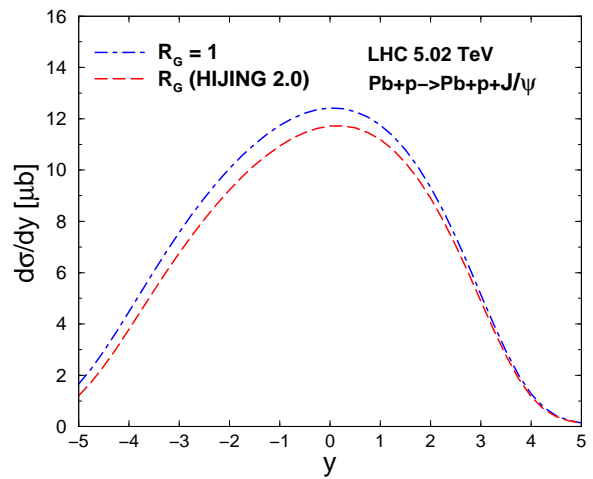}
\caption{(Color online) Rapidity distribution of $\psi(1S)$ production in $pA$ collisions at 5.02 TeV (curves with same notation as in Fig. 1).}\label{fig:2}
\end{figure}

\begin{figure}[t]
\includegraphics[scale=0.47]{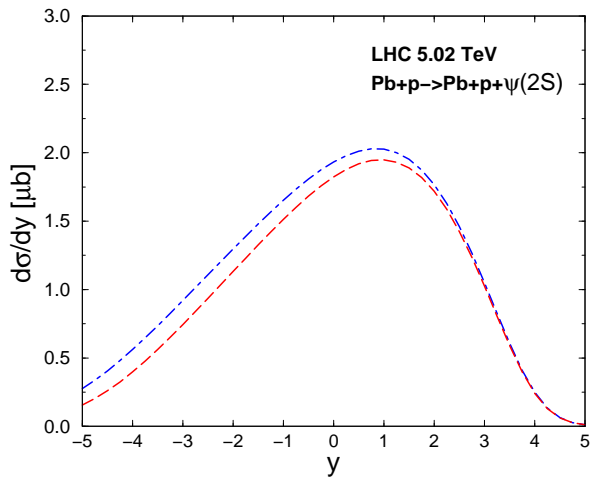}
\caption{(Color online) Rapidity distribution of $\psi(2S)$ production in $pA$ collisions at 5.02 TeV (curves with same notation as in Fig. 1)}\label{fig:3}
\end{figure}

In Ref. \cite{GGM}, the gluon shadowing correction was introduced and the model sensitivity was analyzed. It was found a strong model dependence on predictions. For our purpose here, we will use the simple parameterization of $R_G(x,b)$ given by $\mathrm{HIJING}$ 2.0 \cite{Hijing} (with $s_g=0.17$), which accounts for the impact parameter dependence of nuclear gluon ratio.  As a cross check, in Fig. \ref{fig:1} we show the results using $R_G=1$ (dot-dashed curve) and the impact parameter gluon ratio (long-dashed curve). The last option is good enough for the remaining discussions. Despite the gluon shadowing to be an issue, the photon-proton contribution dominates as the photon flux emitted by the nucleus is enhanced compared to the one coming from proton and the corresponding nuclear suppression of $\sigma_{\gamma A\rightarrow VA}$. For lead-proton collisions at 5.02 TeV and rapidity $y=-3$ the quarkonium production on the proton (low energy photons from nucleus) probes Bjorken variable values $x_p\simeq 10^{-2}$ whereas the production on the nucleus  (high energy photons from proton) probes values $x_A\simeq 10^{-5}$. In Fig. \ref{fig:2} we present the results for the rapidity distribution of $\psi(1S)$ state in $Pb+p$ collisions at $\sqrt{s_{pA}}=5.02$ TeV. The dot-dashed curve is obtained using $R_G=1$ and the long-dashed one is for the $R_G(x,b)$ from $\mathrm{HIJING}$ 2.0. As expected the gluon shadowing effect has a small impact in $pA$ case as the photon-proton interaction is a minor contribution and it is relevant in the negative rapidity region (small $x_A$). We have checked that the rapidity distribution is suppressed by a factor 0.72 at $y=-5$ to 0.94 at $y=0$. In addition,  in Fig. \ref{fig:3} we present the results for the excited state $\psi (2S)$, using the same notation for the figure labels. The suppression is similar to the $J/\psi$ case, being a factor 0.62 at $y=-5$ as expected due the node effects (more shadowing compared to the fundamental state). 

\begin{figure}[t]
\includegraphics[scale=0.47]{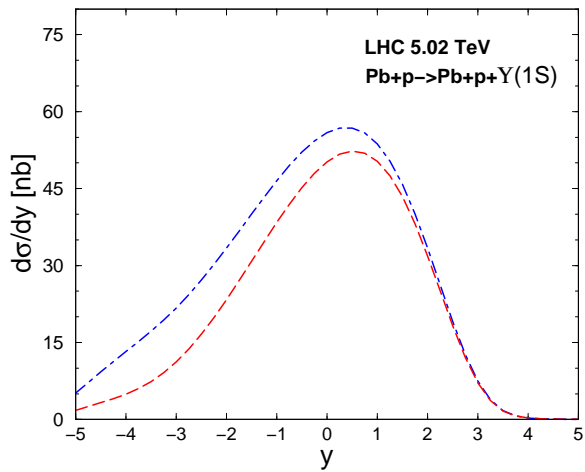}
\caption{(Color online) Rapidity distribution of $\Upsilon$ production in $pA$ collisions at 5.02 TeV (curves with same notation as in Fig. 1).}\label{fig:4}
\end{figure}

Finally, in Fig. \ref{fig:4} the investigation for the $\Upsilon (1S)$ state is done. Clearly, the suppression in negative rapidities is stronger than  for charmonia, mostly for $y<-2$. This is due to the fact that the photon-nucleus interaction contributes more in the $\Upsilon$ case compared to charmonia. Namely, the eikonal giving the multiple scattering of color dipoles is less suppressed compared to $J/\psi$ production. That is directly related to the higher $x_A$ probed in $\Upsilon$ production. For instance, at $y=-3$ one gets $x_A\simeq  10^{-4}$  that can be compared to values $x_A\simeq 10^{-5}$ for charmonia at the same rapidity value.

\begin{figure}[t]
\includegraphics[scale=0.47]{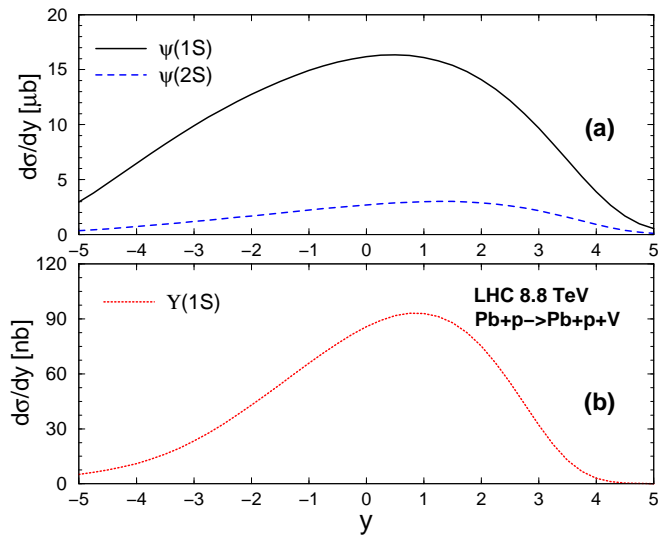}
\caption{(Color online) (a) The rapidity distribution for $\psi(1S)$ (solid line) and $\psi(2S)$ production (dashed line) at energy $\sqrt{s_{pA}}=8.8$ TeV. (b) The rapidity distribution for $\Upsilon(1S)$ production  at energy $\sqrt{s_{pA}}=8.8$ TeV.}
\label{fig:5}
\end{figure}

The predictions for the higher energy $pA$ collisions are presented in Fig. \ref{fig:5}, taking the designed energy of $\sqrt{s_{pA}}=8.8$ TeV. In both plots, we are using $R_G(x,b)$ from $\mathrm{HIJING}$ 2.0. In Fig. \ref{fig:5}-a  we analyze the charmonia production. The solid curve represents the result for $J/\psi$ production, whereas the prediction for $\psi(2S)$ is given by the  dashed curve. The ratio $\psi(2S)/\psi(1S)$ follows the original trend at 5.02 TeV and the cross section is higher by a factor 1.3. In Fig. \ref{fig:5}-b, the prediction for $\Upsilon$ is done. The enhancement in cross section normalization is now a factor 1.5.

Let us now to compare the present results to another approaches. In Ref. \cite{GZ} the $J/\psi$ production in $pA$ has been computed in 5.02 TeV using the LO pQCD calculations. The prediction for the rapidity distribution for $Pb+p\rightarrow Pb+p+J/\psi$ is qualitatively similar to ours. The main difference is in the  very forward negative rapidities, where the distribution is smaller than in \cite{GZ}. The reason is the distinct way to parameterize the threshold $x_g\rightarrow 1$ effect in the photon-proton cross section. A similar LO pQCD calculation has been done also in Ref. \cite{AN} for the $J/\psi$ and $\Upsilon$ production. Our predictions are similar to results using the EKS08 nuclear PDF in \cite{AN}, despite we have not a second peak for $\Upsilon$. The reason should be a stronger suppression for heavier mesons in our case (the $\mathrm{HIJING}$ 2.0  gluon  ratio gives a very strong shadowing effect).  When compared to approaches using color dipole framework, the additional information here is the introduction of gluon shadowing correction which is not so intense as in the PbPb case at the LHC energies. In Ref. \cite{LM}, the production of $J/\psi$ has been computed in $pA$ collisions and our results are similar except in the large negative rapidity region. This is due to our $x_p$-threshold factor imposed to the dipole cross section.  Compared to the previous $pA$ calculation in the color dipole framework \cite{GMpa}, the current study introduces the additional contribution from photon-nucleus interaction.

Finally, let us comment on the theoretical uncertainties associated to the calculations. As pointed out in Ref. \cite{GZ}, the photon flux for protons introduces some uncertainty mostly at large rapidities whereas deviations in modeling the flux for nucleus is considerably smaller. Concerning the color dipole framework considered to compute the photon level cross sections, the main uncertainties come from the model for the meson wavefunction and on the choice for the dipole-proton cross section. The uncertainty on modeling the wavefunction gives a uncertainty of order 13 \% (for fixed dipole-cross section), whereas one has 5 \% uncertainty on modeling the dipole cross section in $pA$ case. We would have a considerable uncertainty on the nuclear shadowing factor $R_G(x,Q^2,b)$, which is very important in the negative rapidity region. We have constrained it by describing the $J/\psi$ cross section in PbPb collisions measured by ALICE. 

\section{Summary}

An investigation was done on the coherent photoproduction of charmonia and $\Upsilon$ in the proton-nucleus collisions in the  LHC energies. It was included both contributions of photon-proton and proton-nucleus interactions. Predictions for the rapidity distributions are presented  using the color dipole formalism and including saturation effects that are expected to be relevant at high energies. It was found that the photon-nucleus contribution is not so important for the charmonium production but relevant for the $\Upsilon$ case. The calculation is consistent with another approaches, as the LO pQCD formalism and previous color dipole calculations. The gluon shadowing effect is considered and the impact is smaller than for the PbPb case. Predictions were made for the next run of LHC considering proton-lead collisions and the main features present at 5.05 TeV are still present.

\begin{acknowledgments}
This work was  partially financed by the Brazilian funding
agency CNPq and by the French-Brazilian scientific cooperation project CAPES-COFECUB 744/12. MVTM thanks the kind hospitality at IPhT CEA Saclay (France), where this work was accomplished.

\end{acknowledgments}


\begin{thebibliography}{99}
\bibitem{nik} N. N. Nikolaev, B. G. Zakharov,  Phys. Lett. B  {\bf 332}, 184 (1994); {Z. Phys. C} {\bf 64}, 631 (1994).

\bibitem{Nemchik:1996pp} 
  J.~Nemchik, N.~N.~Nikolaev, E.~Predazzi and B.~G.~Zakharov,
  %``Color dipole systematics of diffractive photoproduction and electroproduction of vector mesons,''
  Phys.\ Lett.\ B {\bf 374}, 199 (1996).

\bibitem{hdqcd} 
  F.~Gelis, E.~Iancu, J.~Jalilian-Marian and R.~Venugopalan,
    Ann.\ Rev.\ Nucl.\ Part.\ Sci.\  {\bf 60}, 463 (2010);
  H.~Weigert,  Prog.\ Part.\ Nucl.\ Phys.\  {\bf 55}, 461 (2005); J.~Jalilian-Marian and Y.~V.~Kovchegov, Prog.\ Part.\ Nucl.\ Phys.\  {\bf 56}, 104 (2006).

\bibitem{upcs}
 G. Baur, K. Hencken, D. Trautmann, S. Sadovsky, Y. Kharlov, Phys.
Rep. {\bf 364}, 359 (2002);
 C.~A. Bertulani, S.~R.~Klein and J.~Nystrand, Ann. Rev. Nucl. Part. Sci. {\bf 55}, 271 (2005).

\bibitem{KN} S. Klein and J. Nystrand, Phys. Rev. {\bf C60}, 014903 (1999).

\bibitem{H1psi2} C. Adloff {\it et al.} [H1 Collaboration], Phys. Lett. B541, 251 (2002).

\bibitem{GGM} M.B. Gay Ducati, M.T. Griep, M.V.T. Machado, Phys. Rev.{\bf D88} 017504 (2013); M.B. Gay Ducati, M.T. Griep, M.V.T. Machado, Phys. Rev. {\bf C 88} 014910 (2013).

\bibitem{Boris} B. Z. Kopeliovich and B. G. Zakharov, Phys. Rev. D 44, 3466 (1991). 

\bibitem{Borispsi} Y.~.P.~Ivanov, B.~Z.~Kopeliovich, A.~V.~Tarasov and J.~Hufner,
  %``Electroproduction of charmonia off nuclei,''
  Phys.\ Rev.\ C {\bf 66}, 024903 (2002).

\bibitem{Sandapen}
J.~R.~Forshaw, R.~Sandapen and G.~Shaw,
  %``Further success of the colour dipole model,''
  JHEP {\bf 0611}, 025 (2006);  B.E. Cox, J.~R.~Forshaw and  R.~Sandapen, JHEP {\bf 0906 }, 034 (2009).

\bibitem{IIM}
  E.~Iancu, K.~Itakura and S.~Munier,
  %``Saturation and BFKL dynamics in the HERA data at small x,''
  Phys.\ Lett.\ B {\bf 590}, 199 (2004).

\bibitem{AmirSchmidt} A.H. Rezaeian and I. Schmidt, Phys. Rev. {\bf C 88}, 074016 (2013).

\bibitem{ALICE1}  B. Abelev {\it et al.} [ALICE Collaboration], Phys. Lett. B718, 1273 (2013).

\bibitem{ALICE2}  E. Abbas {\it et al.} [ALICE Collaboration], Eur. Phys. J. {\bf C73}, 2617 (2013).

\bibitem{KST}  B.~Z.~Kopeliovich, A. Schaefer and A.~V.~Tarasov,
  Phys.\ Rev.\ C {\bf 62}, 054022 (2000).

\bibitem{Hijing} W.-T. Deng, X.-N. Wang and R. Xu, Phys. Rev. {\bf C 83}, 014915 (2011).

\bibitem{GZ} 
   V.~Guzey and M. Zhalov, arXiv:1307.6689 [hep-ph].

\bibitem{AN} A. Adeluyi and A. Nguyen, Phys. Rev. C {\bf 87}, 027901 (2013).

\bibitem{LM} 
  T.~Lappi and H.~Mantysaari,
  %``J/Psi production in ultraperipheral Pb+Pb and p+Pb collisions at LHC energies,''
Phys. Rev. C {\bf 87}, 032201 (2013). 

\bibitem{GMpa}  V.P. Gon\c{c}alves and M.V.T. Machado, Phys. Rev. {\bf C 73}, 044902 (2006).

\end{thebibliography}
\end{document}